\documentclass[aps,twocolumn,preprintnumbers,amsmath,amssymb,superscriptaddress,showpacs]{revtex4-2}

\usepackage{graphicx}
\usepackage{epstopdf}
\usepackage{amsmath}
\usepackage{multirow}
\usepackage{xcolor}
\usepackage{ulem}
\usepackage[version=4]{mhchem}
\usepackage{hyperref}

\hypersetup{
    colorlinks=true, 
    linkcolor=blue,  
    citecolor=blue,   
    filecolor=magenta, 
    urlcolor=blue,   
}
\newcommand{\RNum}[1]{\uppercase\expandafter{\romannumeral #1\relax}}

\begin{document}

\title{Bridging Crystal Structure and Material Properties via Bond-Centric Descriptors}

\author{Jian-Feng Zhang}\email{jianfeng.zhang@hpstar.ac.cn}
\affiliation{ Center for High Pressure Science and Technology Advanced Research, Beijing 100193, China. }

\author{Ze-Feng Gao}
\affiliation{ School of Physics and Beijing Key Laboratory of Opto-electronic Functional Materials \& Micro-nano Devices, Renmin University of China, Beijing 100872, China }
\affiliation{ Key Laboratory of Quantum State Construction and Manipulation (Ministry of Education), Renmin University of China, Beijing 100872, China }

\author{Xiao-Qi Han}
\affiliation{ School of Physics and Beijing Key Laboratory of Opto-electronic Functional Materials \& Micro-nano Devices, Renmin University of China, Beijing 100872, China }
\affiliation{ Key Laboratory of Quantum State Construction and Manipulation (Ministry of Education), Renmin University of China, Beijing 100872, China }

\author{Bo Zhan}
\affiliation{ Institute of Physics, Chinese Academy of Sciences, Beijing 100190, China}

\author{Dingshun Lv}
\affiliation{ Field Quantum Research Institute, Beijing 100084, China }

\author{Miao Gao}
\affiliation{Department of Physics, School of Physical Science and Technology, Ningbo University, Zhejiang 315211, China}

\author{Kai Liu}\email{kliu@ruc.edu.cn} 
\affiliation{ School of Physics and Beijing Key Laboratory of Opto-electronic Functional Materials \& Micro-nano Devices, Renmin University of China, Beijing 100872, China }
\affiliation{ Key Laboratory of Quantum State Construction and Manipulation (Ministry of Education), Renmin University of China, Beijing 100872, China }

\author{Xinguo Ren}\email{renxg@iphy.ac.cn}
\affiliation{ Institute of Physics, Chinese Academy of Sciences, Beijing 100190, China}

\author{Zhong-Yi Lu}\email{zlu@ruc.edu.cn} 
\affiliation{ School of Physics and Beijing Key Laboratory of Opto-electronic Functional Materials \& Micro-nano Devices, Renmin University of China, Beijing 100872, China }
\affiliation{ Key Laboratory of Quantum State Construction and Manipulation (Ministry of Education), Renmin University of China, Beijing 100872, China }

\author{Tao Xiang}\email{txiang@iphy.ac.cn}
\affiliation{ Institute of Physics, Chinese Academy of Sciences, Beijing 100190, China}
\affiliation{ School of Physical Sciences, University of Chinese Academy of Sciences, Beijing 100049, China}

\date{\today}

\begin{abstract}

Although chemical bonding is the fundamental mechanistic bridge connecting atomic structure to macroscopic material properties, current data-driven materials science largely treats it as an implicit ``black box". Existing machine-learning (ML) models rely predominantly on geometric coordinates, forcing them to infer complex quantum-mechanical relationships from scratch. This lack of intermediate physical features limits model interpretability and generalizability, particularly when training data are scarce. To solve this problem, we introduce MattKeyBond, a bond-centric materials database that explicitly maps the local electronic landscape and bonding interactions of materials. Building on this, we propose Bonding Attractivity (BA), a novel element-specific descriptor that quantifies the intrinsic capability of atoms to form covalent networks. By providing precomputed, energy-resolved bonding descriptors, MattKeyBond transforms the implicit ``black box" into a set of physically interpretable features. This strategy relieves ML models of the burden of deducing physical laws from pure geometry, enabling accurate predictions even with limited data and seamlessly integrating electronic structure theory into modern AI workflows.

\end{abstract}


\maketitle

\section{Introduction}

Crystal structure dictates the physical properties of materials, yet it is chemical bonding that mechanistically links atomic arrangement to electronic structure and macroscopic behavior. In solids, each atom interacts with a local environment determined by neighboring atoms and the resulting crystal field. Through field-induced charge redistribution and orbital hybridization, inter-atomic bonding yields an inherent electronic structure strictly dependent on atomic arrangement. Atom- and bond-resolved analyses are therefore essential for establishing mechanistic connections between structure and properties, and for enabling the rapid screening and design of materials.

\begin{figure}[t]
\includegraphics[angle=0,scale=0.38]{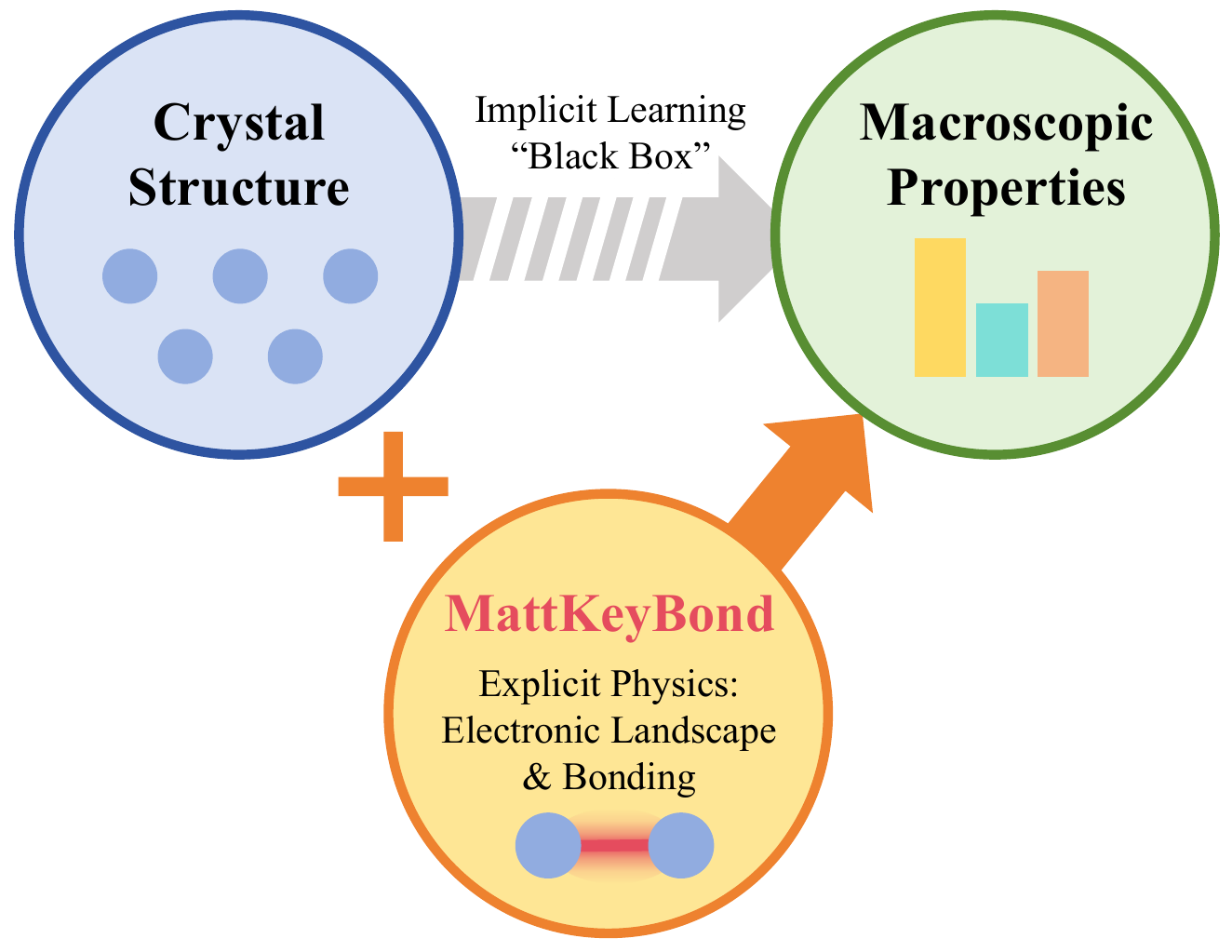}
\caption{
 MattKeyBond: Enhancing implicit learning with interpretable electronic and bonding descriptors.
 }
\label{Fig_outline}
\end{figure}

However, despite the emergence of a rich ecosystem of materials databases, current resources generally lack this crucial information. Existing platforms primarily capture structural geometry and global scalar properties. General-purpose databases, such as the Materials Project (MP)\cite{mp}, Atomly\cite{atomly}, OQMD\cite{oqmd1,oqmd2}, AFLOW\cite{aflow}, and NOMAD\cite{nomad}, aggregate high-throughput calculations to provide structures, formation energies, and electronic band structures. Similarly, domain-targeted databases like C2DB\cite{c2db1,c2db2}, JARVIS-DFT\cite{jarvis}, CoRE MOF\cite{coremof}, and the Open Catalyst Project (OC20/OC22)\cite{oc20,oc22} focus on specific material classes. Despite their breadth, the absence of physics-based intermediate features limits the effectiveness of current data-driven materials science. Models are therefore forced to rely predominantly on geometric information \cite{CGCNN,MEGNET,ALIGNN,CHGNET}, effectively treating the chemical bond as an implicit ``black box" (top panel of Figure \ref{Fig_outline}). This compels machine-learning models to implicitly relearn complex quantum-mechanical relationships from geometry alone, which limits their interpretability and generalizability, particularly in complex systems where experimental data are scarce, such as superconductors. 

While recent efforts have attempted to incorporate electronic features by learning continuous charge densities or DFT Hamiltonians \cite{deepH,deepC}, a comprehensive, bond-centric database remains absent.
To bridge this gap and decode the underlying interactions, we introduce \textbf{MattKeyBond}, a bond-centric materials database constructed from high-throughput first-principles calculations. Leveraging the Closest Wannier Functions (CWF)\cite{cwf} method and integral crystal orbital Hamilton population (ICOHP) analyses\cite{cohp0,cohp,cohp2}, MattKeyBond transcends conventional structural geometry to map the electronic landscape into a real-space, bond-resolved representation. Through its high-fidelity, atom-pair-resolved features, including charge transfer, orbital Hamiltonians, bond energy, and bond-order-related density matrices, MattKeyBond enables systematic, searchable, and comparable bonding analysis across 36,377 inorganic compounds in the current release. 

Building on this extensive dataset, we further propose \textbf{Bonding Attractivity (BA)}, a novel element-specific descriptor designed to quantify the intrinsic capability of atoms to form covalent networks. While traditional electronegativity and its modern extensions\cite{en,en2} primarily characterize the tendency for charge transfer (ionicity), they do not fully capture the energetic contributions of shared electrons (covalency). BA complements this classical concept by explicitly measuring the strength of orbital hybridization that binds atoms together. Leveraging 3.6 million bond records, we have successfully parametrized BA for elements from hydrogen ($Z=1$) to bismuth ($Z=83$), quantifying its dependence on bond length and valence state. As a human-readable descriptor, this metric provides an intuitive way to understand the intrinsic bonding ability of elements in covalent systems.

Crucially, we position MattKeyBond and BA not merely as static data, but as physically interpretable intermediate features that connect atomic coordinates to macroscopic behavior. Unlike purely structural inputs, these descriptors add an explicit energy dimension derived from rigorous electronic structure theory. For the ``AI for Science" community,  this work represents a clear strategy: by providing precomputed quantum insights, we relieve machine-learning models of the burden of learning physical laws from scratch. MattKeyBond thus serves as a foundational resource, integrating electronic structure theory into modern AI workflows to accelerate bonding-guided materials discovery.

\section{Descriptor of Chemical bond}

\begin{figure*}[t]
\includegraphics[angle=0,scale=0.5]{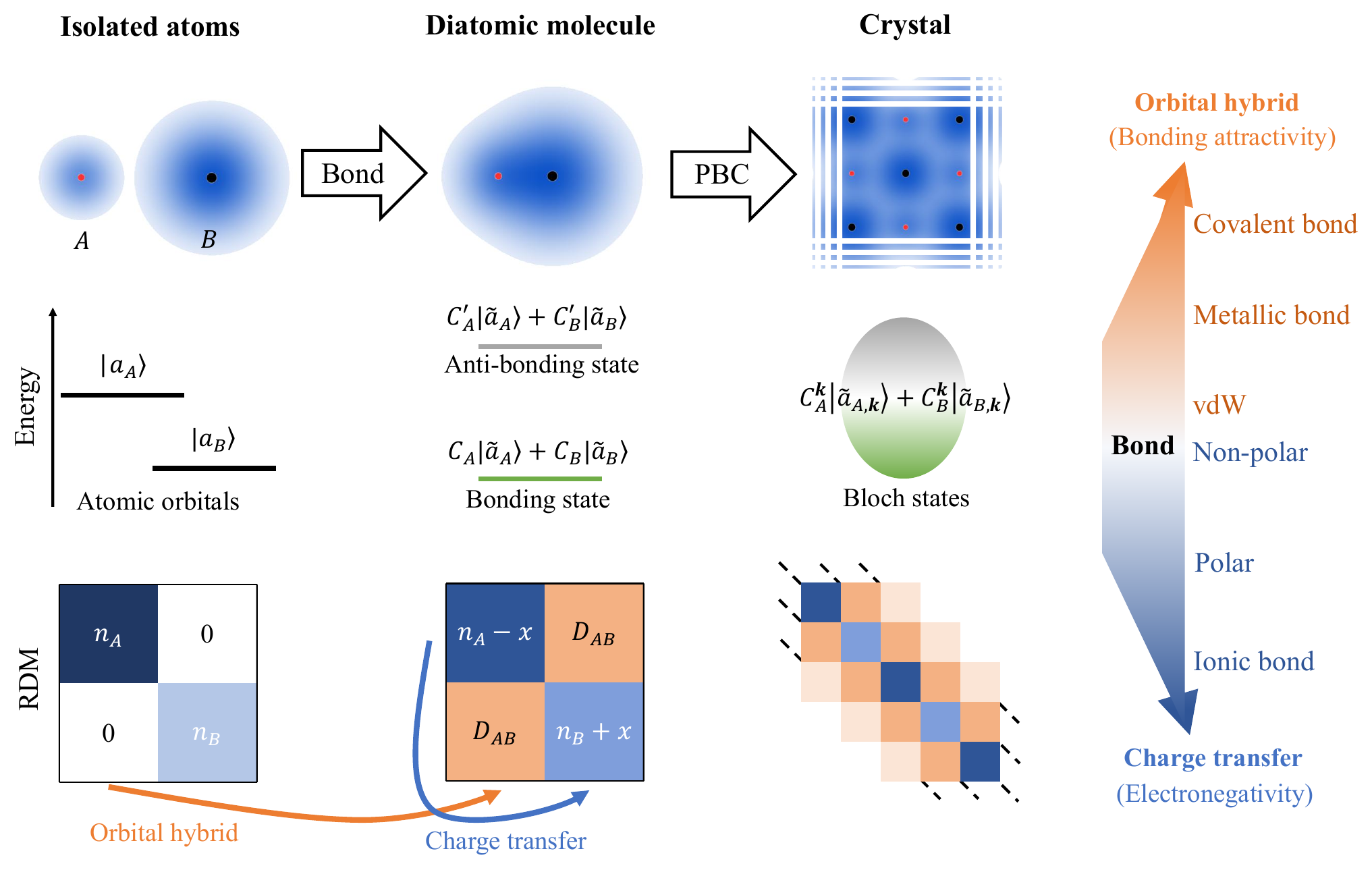}
\caption{
 The evolution of electronic structure from isolated atomic orbitals to hybridized molecular orbitals and finally to Bloch states in a periodic crystal. The reconstruction of the reduced density matrix (RDM) highlights two distinct energetic pathways driving bond formation: charge transfer (represented by blue arrows), and orbital hybridization (represented by orange arrows). The last column lists a conceptual classification of chemical bonds based on these two primary dimensions.
 }
\label{Fig_sum}
\end{figure*}

Fundamentally, material synthesizability is governed by energy minimization, a process intrinsically driven by the formation of chemical bonds. The release of energy serves as a quantitative measure of bond strength, typically described by metrics such as cohesive energy, bond energy, formation energy, or energy above the hull (depending on the different reference used). Consequently, beyond specific electronic or mechanical properties, a numerical descriptor of chemical bond strength offers a direct metric for energetic stability. Below, we briefly outline the formation of chemical bonds, their descriptors, and their relationship to bond energy. 

As illustrated in Figure \ref{Fig_sum}, consider isolated atoms $A$ and $B$. Their atomic orbitals, denoted as $|a_{A/B}\rangle$, exhibit element-specific spherical harmonic distributions and energy levels (omitting orbital indices for simplicity). Taking a simple diatomic molecule as an example, as atoms $A$ and $B$ approach each other, their orbitals undergo overlap, renormalization, and hybridization under the influence of the local field, resulting in new molecular orbitals $|n\rangle$:
\begin{equation}
\label{eq_sps}
|n\rangle=C_A^n\left|\tilde{a}_A\right\rangle+C_B^n\left|\tilde{a}_B\right\rangle. 
\end{equation}
Here, $|\tilde{a}_{A/B}\rangle$ represents the renormalized atomic orbital in the local environment, and $C_{A/B}^n$ denotes the combination coefficients. Within the framework of density functional theory (DFT)\cite{dft1,dft2}, this process is described by the redistribution of charge density and self-consistent field. In the single-particle representation, this redistribution is reflected in the reduced density matrix (RDM) of $\langle\tilde{a}_A |\hat{D}|\tilde{a}_B \rangle$. As illustrated in the third row of Fig. \ref{Fig_sum}, along the bond between $A$ and $B$, the RDM is reconstructed in two primary ways: one is the hybridization between atomic orbitals (orange arrows), which provides the cohesive force between the bonding atoms; the other is the charge transfer from one atom to another (blue arrows), resulting in their respective valence states $x$. These two aspects represent two major pathways for energy release during bond formation.

Within the DFT framework\cite{dft1,dft2}, the total energy of a system can be decomposed into a band-structure term, a double counting correction to the Hartree and exchange-correlation energies, and the ion-ion Ewald energy:
\begin{equation}
E_{\mathrm{tot}} = E_{\mathrm{band}} - E_{\mathrm{dc}} + E_{\mathrm{Ewald}}.
\end{equation}
The $E_{\mathrm{Ewald}}$ term arises from ion-ion Coulomb repulsion, which generally increases as bonding atoms approach each other. $E_{\mathrm{dc}}$ strongly depends on the charge density derived from the eigenstates $|n\rangle$ of $E_{\mathrm{band}}$, and generally follows a trend similar to that of $E_{\mathrm{band}}$ due to the self-consistent nature of the calculation. Consequently, during chemical bond formation, the energy gain primarily originates from the first term, $E_{\mathrm{band}}$, which sums the energies $\varepsilon_n$ of all occupied single-particle eigenstates $|n\rangle$. Within the subspace of the atomic-orbital basis, $E_{\text{band}}$ can be further separated into an on-site term and an inter-site hybridization term:
\begin{equation}
E_{\mathrm{band}} = E_{\mathrm{onsite}} + E_{\mathrm{hybrid}},
\end{equation}
which respectively correspond to the two major processes in bond formation: charge transfer and orbital hybridization.

Furthermore, the respective contributions from $E_{\mathrm{onsite}}$ and $E_{\mathrm{hybrid}}$ provide a basis for classifying chemical bonds. As illustrated in the last column of Fig. \ref{Fig_sum}, based on the tendency for charge transfer, bonds can be described as non-polar (no transfer), polar (partial transfer), or ionic (full transfer). From the perspective of orbital hybridization strength, one can roughly distinguish between the weakest van der Waals (vdW) interactions, weak metallic bonds, and the strongest covalent bonds.

For any energy $\varepsilon_n$ contributing to $E_{\text{band}}$, COHP analysis resolves it into pairwise contributions\cite{cohp0,cohp,cohp2}:
\begin{equation}
\varepsilon_{n}=\sum_{A,B}\text{COHP}_{A,B}^n=\text{Re}\sum_{A,B}C^{n*}_AC^n_B \langle\tilde{a}_A |\hat{H}|\tilde{a}_B \rangle.
\end{equation}
The sign of $\text{COHP}_{AB}^n$ characterizes the nature of the interaction: $\text{COHP} < 0$ indicates a bonding state, while $\text{COHP} > 0$ indicates an anti-bonding state between atoms A and B. Additionally, based on the magnetic quantum number $m$ of $|a_{A/B}\rangle$ (aligning the $A$-$B$ bond along the $z$-axis), they can hybridize into strong $\sigma$ bonds ($m = 0$), intermediate $\pi$ bonds ($m = \pm 1$), or weaker $\delta$ bonds ($m = \pm 2$).

Summing over all occupied states, the integral of COHP (ICOHP) yields:
\begin{equation}
\begin{gathered}
E_{\mathrm{onsite}} = \sum_{A} \mathrm{ICOHP}_{AA},\\
E_{\mathrm{hybrid}} = \sum_{A\ne B} \mathrm{ICOHP}_{AB},
\end{gathered}
\end{equation}
which enables bond-resolved energetics for every inter-atomic pair $(A, B)$ in complex structures. 
This energy integration also simplifies the calculation of $\mathrm{ICOHP}_{AB}$:
\begin{equation}
\mathrm{ICOHP}_{AB}=\langle\tilde{a}_B |\hat{D}|\tilde{a}_A \rangle\langle\tilde{a}_A |\hat{H}|\tilde{a}_B \rangle
\end{equation}
Generally, $\text{ICOHP} < 0$ indicates a stabilizing interaction. For crystal systems (third column of Fig. \ref{Fig_sum}) with periodic boundary conditions (PBC), the combination coefficients in Eq.~\ref{eq_sps} acquire a Bloch phase factor in the momentum-$\boldsymbol{k}$ representation. While discrete molecular orbital levels evolve into continuous band structures, the physical basis of the chemical bond and the analytical tools of ICOHP remain identical to those used in molecular systems. See the Appendix B and C for more details regarding PBC and multi-orbital systems. 

Although ICOHP does not account for all contributions to the total bond energy, it provides a direct measure of the energy associated with orbital hybridization between specific atomic pairs. Furthermore, this hybridization energy is intrinsically tied to inter-atomic binding forces and thus informs a range of material properties, including hardness, inter-atomic force constants (phonon stiffness), and, in metal systems, electron-phonon coupling strength, electronic transport, and, superconductivity\cite{mgb2}. Therefore, in this work, we adopt ICOHP as a primary descriptor of bond strength for the construction of the database.

\section{High-Throughput workflow}

\begin{figure}[t]
\includegraphics[angle=0,scale=0.4]{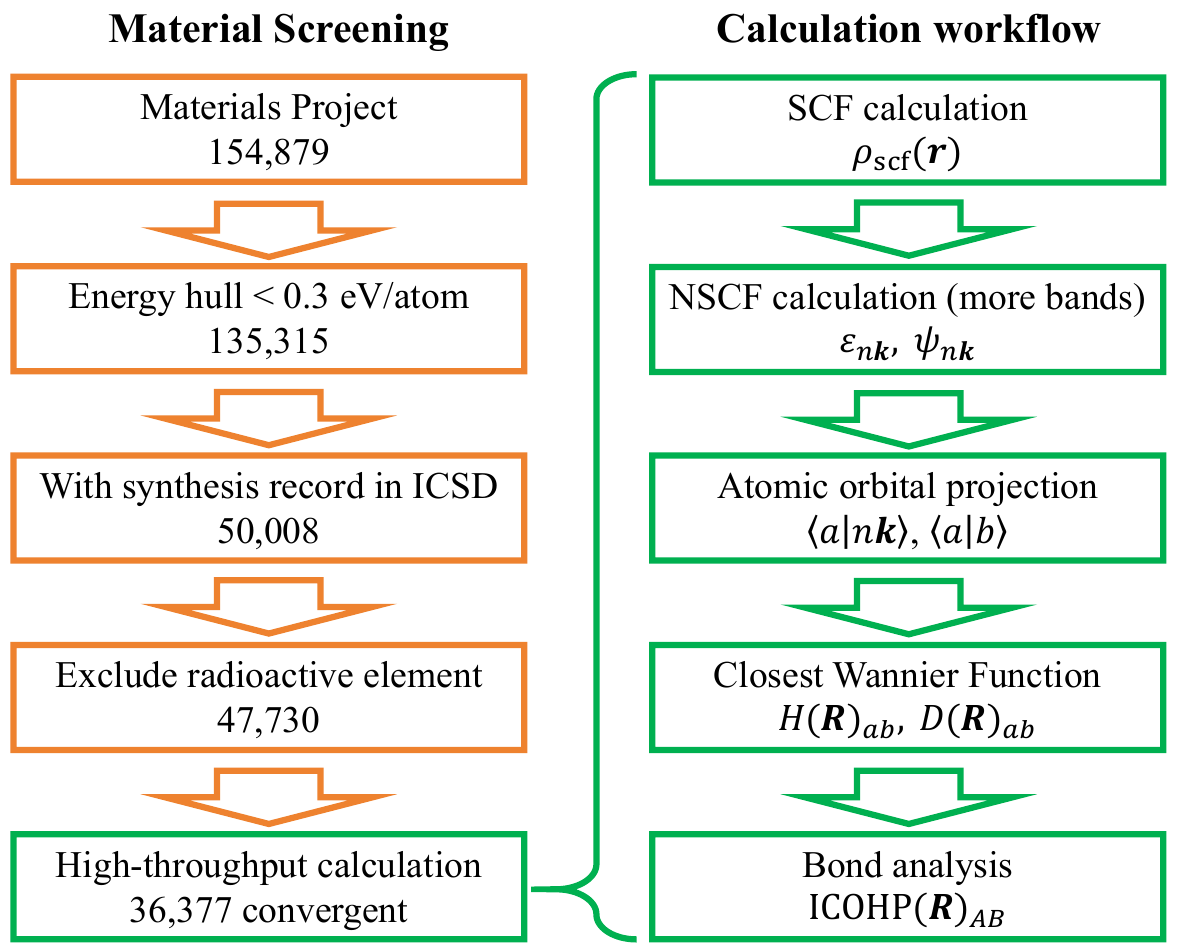}
\caption{
Materials-screening criteria and the high-throughput calculation/analysis workflow. }
\label{Fig_wf}
\end{figure}

\begin{figure*}[t]
\includegraphics[angle=0,scale=0.55]{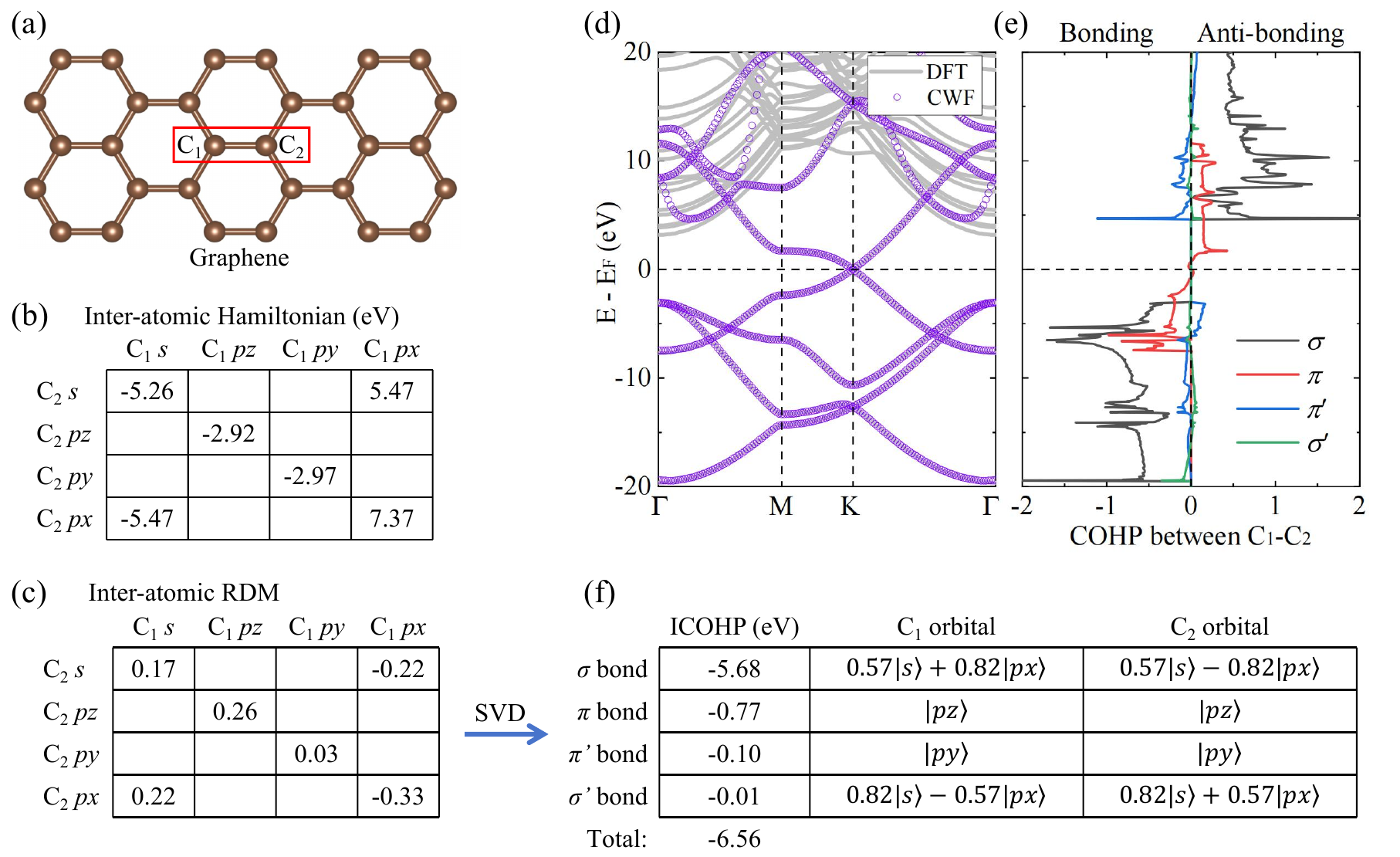}
\caption{
Detailed bonding analysis of the nearest C-C bond in graphene as a representative example. (a) Crystal structure of graphene with the target C$_1$-C$_2$ atomic pair highlighted. (b) The real-space inter-atomic Hamiltonian matrix and (c) RDM for the C$_1$-C$_2$ pair in the atomic orbital basis ($s, p_z, p_y, p_x$). (d) Comparison between the electronic band structure calculated by DFT (gray lines) and that obtained from our CWF model (purple circles). (e) Bond-resolved COHP curves in energy space, clearly distinguishing the contributions and strengths from $\sigma$ (black), $\pi$ (red), $\pi'$ (blue), and $\sigma'$ (green). (f) Quantitative decomposition of the bond strength via the SVD of RDM. 
 }
\label{Fig_example}
\end{figure*}

In this work, we first retrieved all crystal structures available in the MP database\cite{mp}, comprising 154{,}879 entries as of October 2025. The detailed screening criteria and the computational workflow are summarized in Fig.~\ref{Fig_wf}. To ensure energetic stability, we retained compounds with an energy above the hull of less than 0.3 eV/atom, resulting in 135{,}315 stable or metastable structures. To further guarantee synthesizability, we restricted the set to compounds with experimental records in the Inorganic Crystal Structure Database (ICSD)\cite{icsd}, yielding 50{,}008 materials. We additionally excluded systems containing radioactive elements, resulting in 47{,}730 candidates for high-throughput calculations. At present, within the constraints of the allotted wall time and convergence criteria, 36{,}377 materials have been successfully computed and are included in this initial release.

For each material, the calculations and analyses proceed in five steps:

 \textbf{1. Self-consistent field (SCF) calculation.}
We performed DFT SCF calculations using the Quantum ESPRESSO (QE) package~\cite{pwscf} to obtain the electronic ground state and charge density. Additional settings (functional\cite{PBE}, pseudopotentials\cite{psl}, $\boldsymbol{k}$-point meshes, smearing, and convergence thresholds) are provided in Appendix A.

 \textbf{2. Non-self-consistent field (NSCF) calculation.}
For subsequent Wannier downfolding calculations, we carried out NSCF calculations with an increased number of bands. The total number of bands was empirically set to
\[
\texttt{nbnd} = N_{\mathrm{atomorb}} + 5N_{\mathrm{atom}}.
\]
Here, $N_{\mathrm{atomorb}}$ is the total number of atomic valence orbitals included in the projection and $N_{\mathrm{atom}}$ is the number of atoms in the primitive cell (see TABLE \ref{tab_ele} in the Appendix). Both SCF and NSCF steps were executed with \texttt{pw.x} in QE\cite{pwscf}.

 \textbf{3. Atomic-orbital projections.}
Using the \texttt{projwfc.x} module of QE\cite{pwscf}, we computed (i) the projection matrix between atomic orbitals and Kohn-Sham eigenstates, $\langle a \mid n \rangle$, and (ii) the overlap integrals between neighboring atomic orbitals, $\langle a \mid b \rangle$. Here $|a \rangle$ and $|b \rangle$ denote atomic-like orbitals, and $|n \rangle$ denotes a Kohn-Sham eigenstate obtained from the DFT calculations above.

 \textbf{4. Closest Wannier Function (CWF) downfolding.}
To capture the orbital renormalization in the crystal field, we constructed CWFs using atomic orbitals as guiding functions to downfold the plane-wave Kohn-Sham subspace into a compact Wannier basis\cite{cwf}. This procedure yields a CWF-based tight-binding Hamiltonian and reduced density matrix (RDM) for each system. Algorithmic details and numerical settings are provided in Appendix B.

 \textbf{5. ICOHP-based bond analysis.}
We analyzed bonding using the crystal orbital Hamilton population (COHP) and its energy integral (ICOHP) for all inter-atomic pairs with real-space distances shorter than 6~\AA. The resulting bond-resolved energetics were used to quantify bond strength across all distinct bonded pairs\cite{cohp0,cohp,cohp2}. Additional details are provided in Appendix C.

To demonstrate the microscopic insights provided by MattKeyBond, we present a representative analysis of the nearest carbon-carbon bond in graphene in Figure \ref{Fig_example}. As shown in panel (d), the band structure obtained from our CWF downfolding (purple circles) accurately reproduces the DFT band structure (gray lines) around the Fermi level. Beyond global electronic structure, MattKeyBond provides the local inter-atomic Hamiltonian in panel (b) and the RDM in panel (c) for every orbital pair. These matrices allow for advanced post-processing, such as decomposing the total interaction into independent bonding channels via the singular value decomposition (SVD) of the inter-atomic RDM block.
As shown in panel (f), this analysis can automatically identify the strong $\sigma$ bond (ICOHP = -5.68 eV) formed by $sp^2$-like hybrids (${\sqrt{1/3}}|s \rangle \pm {\sqrt{2/3}} |p_x \rangle$) and the characteristic $\pi$ bond (ICOHP = -0.77 eV) formed by vertical $p_z$ orbitals. 
The remaining two are the in-plane $\pi'$ bond formed by $p_y$ orbitals and the $\sigma'$ bond resulting from the tail-to-tail overlap of $|s \rangle$ and $|p_x \rangle$ orbitals. Together with their COHP distributions in energy space shown in panel (e), these analyses allow users to go beyond total bond energies and dissect the specific orbital mechanisms driving material properties (e.g. high-\(T_c\) superconductivity in MgB\(_2\), driven by metallized B-B \(\sigma\) bonds \cite{mgb2}). 

\begin{figure*}[t]
\includegraphics[angle=0,scale=0.5]{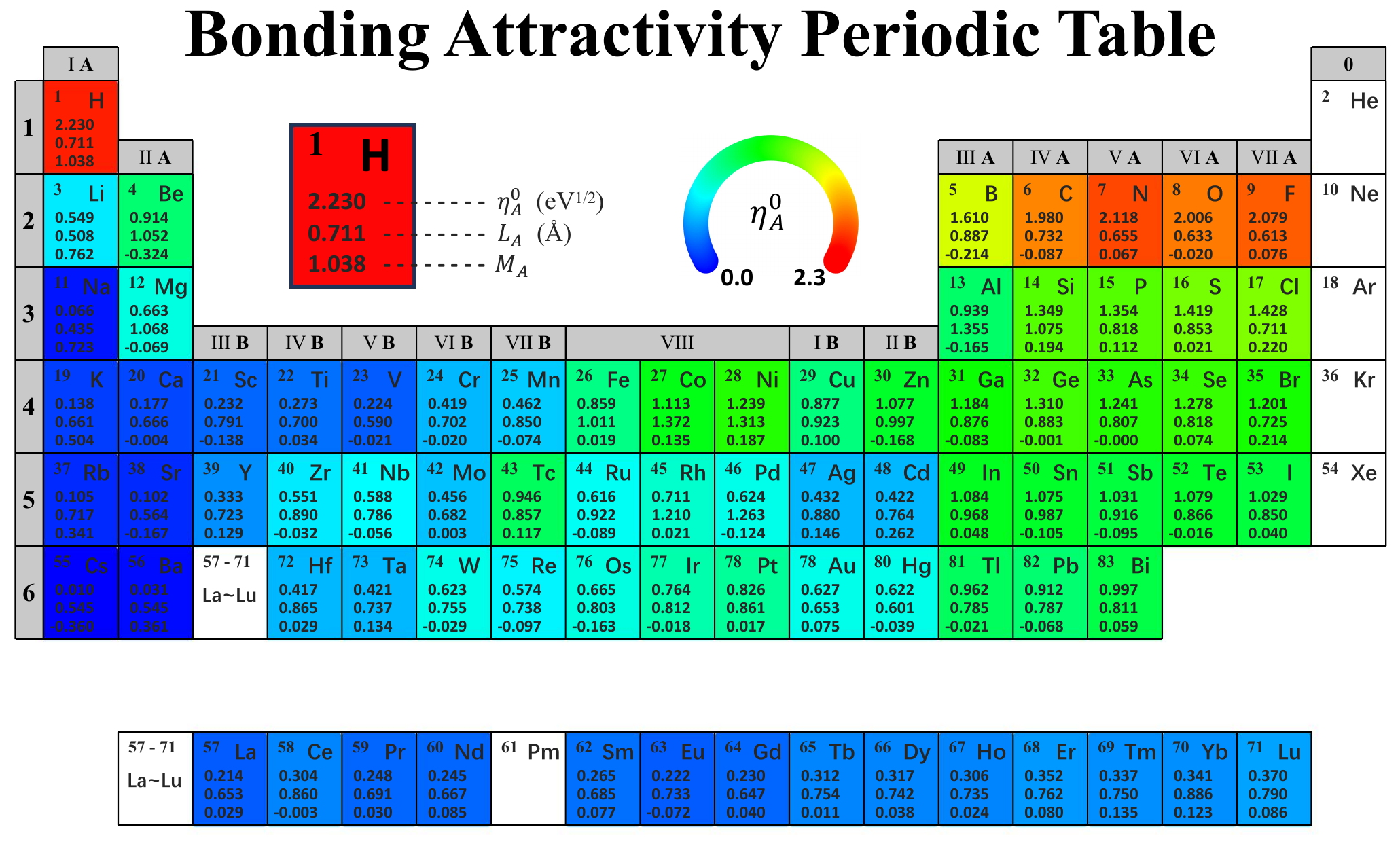}
\caption{
Periodic table of Bonding Attractivity. The three numbers in each element block are, respectively, the intrinsic bonding attractivity $\eta_A^{0}$, the characteristic decay length $L_A$, and the valence-state modulation factor $M_A$. The background color encodes $\eta_A^{0}$.
}
\label{Fig_BA}
\end{figure*}

\section{Bonding Attractivity}

Element-specific bonding trends are traditionally summarized by electronegativity (EN)\cite{en,en2}, such as that represented by the Pauling scale, which was inferred primarily from experimental bond energies across compounds. 
EN captures an element’s tendency to attract electrons. 
Electrons tend to transfer from elements with low EN to those with high EN. 
Inspired by the definition of Pauling EN, using the 3,665,789 bond records (ICOHP values) currently available across 36,377 materials in MattKeyBond, we introduce Bonding Attractivity (BA), denoted as $\eta_A(R,x_A)$. 
Unlike EN, which primarily characterizes the tendency for charge transfer (ionicity), BA complements this classical concept by quantifying the intrinsic capability of an atom to bind with neighbors via covalent-like orbital hybridization (covalency). 
As discussed previously, orbital hybridization plays a decisive role in many physical properties, including inter-atomic force constants, atomic dimerization, and even electron-phonon coupling strength, which can influence electronic transport and superconductivity. 
Therefore, in addition to providing new insights into structure-bonding relationships, the introduction of BA also enables interpretable property prediction and supports data-driven discovery of functional materials. 

In complex crystal environments, bond strength is influenced by numerous factors, including element identity, bond type, hybridized orbital character, valence state, bond length, and crystal-field effects. To obtain a simple, human-readable, yet tractable descriptor, we neglect the dependence on orbital character and local crystal field but retain bond length $R$ and valence state $x_A$ as two dominant variables controlling the BA of atom $A$.

Analogous to Pauling’s treatment of bond energies\cite{en}, we postulate that the ICOHP between two bonded atoms $A$ and $B$ can be expressed as the product of their respective BAs:
\begin{equation}
-\mathrm{ICOHP}_{AB}(R,x_A,x_B) \;=\; \,\eta_A(R,x_A)\,\eta_B(R,x_B).
\label{eq_icohp}
\end{equation}
Here $\eta(R,x) > 0$ has units of eV$^{1/2}$. To capture bond-length decay and valence-state modulation, we adopt the exponential form
\begin{equation}
\eta_A(R,x_A) \;=\; \eta_A^{0}\,\exp\!\bigl[-(R-2r_A)/L_A + M_A x_A \bigr],
\label{eq_eta}
\end{equation}
with the reference value
\begin{equation}
\eta_A^{0} \;=\; \eta_A(2r_A,0).
\label{eq_eta0}
\end{equation}
Here $\eta_A^{0}$ is an element-specific baseline BA; $r_A$ is the covalent radius of element $A$~\cite{RA}, serving as a reference for typical bond lengths; $L_A$ (in \AA) is the characteristic decay length of BA with respect to $R$; and $M_A$ (dimensionless) quantifies the modulation by the valence state. Consequently, the triplet $(\eta_A^{0}, L_A, M_A)$ fully parameterizes the BA for element $A$. As implied by Eq.~\eqref{eq_eta0}, $({\eta_A^{0}})^2$ corresponds to the $-$ICOHP of an $A$-$A$ bond at $R = 2r_A$ and $x_A=0$, so we treat $\eta_A^{0}$ as the fundamental BA quantity for element $A$. 

Using 3{,}665{,}789 bond records (ICOHP values) spanning diverse element combinations, bond lengths, and valence states collected from 36{,}377 materials in MattKeyBond, we performed a least-squares fit of Eqs.~\eqref{eq_icohp} and \eqref{eq_eta}. The quality of the fit is shown in Figs. \ref{BA-1}$\sim$\ref{BA-4} of the Appendix. Fitted parameters from H (Z = 1) to Bi (Z = 83) are summarized in the periodic-table map of Fig.~\ref{Fig_BA}. In each element block, the three numbers report $\eta_A^{0}$, $L_A$, and $M_A$, and the background color encodes $\eta_A^{0}$ from 0.0 (blue) to 2.3 (red). 

Overall, the distribution of $\eta_A^{0}$ shows broad similarity to classic Pauling EN~\cite{en}, with the following trends:

\begin{enumerate}
\item Elements with valence $2p$ orbitals (B$\sim$F) exhibit high $\eta_A^{0}$, consistent with their propensity to form strong covalent bonds.
\item Elements on the left side of the table: alkali metals, alkaline-earth metals, lanthanides, and many group III–VII {\bf B} elements show relatively low $\eta_A^{0}$, aligning with their common cationic roles in compounds.
\item Heavier $p$-block elements (n $>$ 2) and numerous late transition metals generally exhibit intermediate $\eta_A^{0}$ values.
\end{enumerate}

A notable difference from the Pauling electronegativity scale is that the largest $\eta_A^0$ is obtained for hydrogen, followed by typical $2p$ elements. In contrast, fluorine has the highest EN on the Pauling scale. This discrepancy reflects the distinct physics captured: BA measures orbital-hybridization strength, whereas EN includes more contributions from charge-transfer energetics. Fluorine’s large EN mainly arises from the charge-transfer term rather than hybridization energy, favoring ionic bonding with comparatively weaker covalent binding. Hydrogen, conversely, readily forms strong bonds via hybridization across diverse chemical environments, a feature consistent with its behavior in hydrogen storage materials.

In addition to the baseline $\eta_A^{0}$, Eq.~\eqref{eq_eta} introduces bond-length and valence-state dependencies through $L_A$ and $M_A$ (see Fig. \ref{Fig_LM} for element-resolved maps). The decay length $L_A$ governs how rapidly BA weakens with increasing $R$. A small $L_A$ manifests in two typical ways: it indicates either a tendency to form short, stiff bonds with large force constants (characteristic of $2p$ elements like C, N, O, and F), or a steep rise in bonding attractivity as atoms approach each other (seen in reactive alkali, alkaline-earth, and halogen elements). In both cases, hybridization-driven bonding is favored primarily with nearest neighbors. Conversely, as shown in Fig. \ref{Fig_LM}, only five elements (Al, Co, Ni, Rh, and Pd) exhibit conspicuously large $L_A$ values, suggesting that their chemical bonding capabilities remain relatively stable across varying local environments.

The valence modulation factor $M_A$ displays a sign oscillation across the periodic table. In Fig.~\ref{Fig_LM}, the sign of $M_A$ is encoded by the background color (red for positive, blue for negative). A positive $M_A$ implies that removing electrons (higher oxidation state) enhances BA, whereas a negative $M_A$ indicates that adding electrons strengthens BA. Classical EN arguments suggest that gaining electrons should reduce an atom’s electron-attracting tendency, pointing toward positive $M_A$. This behavior indeed appears for many simple alkali and halogen elements with a single half-filled valence orbital. However, for elements in the middle of the table with multi-orbital valence manifolds, adding an electron may create a new unpaired or half-filled orbital and thereby enhance the hybridization with other atoms, yielding negative $M_A$. Consequently, depending on detailed valence-electron counts and orbital occupations, $M_A$ exhibits oscillatory trends within a period.

We emphasize that the current calculations and analyses are performed exclusively within a nonmagnetic framework.
For systems with robust magnetic order or high Neel/Curie temperatures, the present BA formulation may yield suboptimal descriptions of bonding energetics. Furthermore, in addition to magnetism, the local environment and specific orbital characters can play decisive roles in ICOHP. For elements with a single dominant valence orbital (e.g., many chalcogens and halogens), Eq.~\ref{eq_icohp} captures ICOHP accurately. However, for elements with multiple bonding orbitals (e.g., boron), stronger environmental and orbital dependencies introduce additional complexity, as evidenced by the parity plots comparing BA-fitted and DFT-calculated $-\mathrm{ICOHP}$ values in Figs. \ref{BA-1}$\sim$\ref{BA-4} of the Appendix. Despite these limitations, our simplified BA formulation successfully captures the essential trends of atomic bonding across the periodic table. This validates BA as a compact, intuitive, and interpretable descriptor that compresses complex electronic structure information into a human-readable format. 

\section{Summary}

In summary, we have developed MattKeyBond, a bond-centric materials database constructed through high-throughput first-principles calculations, Closest Wannier Function (CWF) downfolding, and integral crystal orbital Hamilton population (ICOHP) analysis. Currently encompassing 36,377 inorganic compounds and over 3.6 million bond records, this database provides a high-fidelity, real-space mapping of local electronic landscapes and bonding interactions. By moving beyond traditional geometric coordinates and global scalar properties, MattKeyBond extracts explicit, energy-resolved intermediate features at the atom-bond resolution, establishing a mechanistic bridge between atomic arrangements and macroscopic material properties. The dataset presented in this paper is openly available at \href{https://doi.org/10.57760/sciencedb.38081}{ScienceDB}.

Building upon this extensive dataset, we introduced Bonding Attractivity (BA), denoted as $\eta_A(R, x_A)$, a novel element-specific descriptor parameterized by a baseline attractivity, a characteristic decay length, and a valence-state modulation factor. While classical electronegativity primarily characterizes the tendency for ionic charge transfer, BA serves as a crucial complementary metric that quantifies the intrinsic capability of atoms to form bonds via covalent orbital hybridization. By distilling complex quantum-mechanical information into an intuitive, human-readable format, BA provides a tractable and physically interpretable metric for bond strength across diverse chemical environments.

The introduction of MattKeyBond and BA represents a significant advancement for the ``AI for Science" community. These precomputed, physics-based descriptors provide machine-learning models with explicit electronic structure information, relieving them of the burden of implicitly reconstructing complex quantum-mechanical relationships from geometric information alone. This integration fundamentally enhances model interpretability and generalizability, particularly in scenarios where experimental training data are scarce, thereby enabling more robust and accurate property predictions.
Furthermore, as generative AI and foundation models continue to evolve, embedding these physics-rich, bond-resolved descriptors into modern workflows holds immense potential for inverse materials design. 

Moving forward, we plan to continuously expand MattKeyBond by incorporating newly predicted and potentially synthesizable crystal structures. Additionally, future iterations of the database will explicitly account for advanced physical interactions, including spin-orbit coupling and magnetism, to provide an even more comprehensive and rigorous description of chemical bonding. 
We anticipate that this foundational resource will significantly accelerate the data-driven discovery and mechanistic understanding of next-generation functional materials, such as unconventional superconductors, advanced catalysts, and novel energy storage systems.

 \begin{acknowledgments}
This work was supported by a project funded by the China Postdoctoral Science Foundation (No. 2022M723355), Chinese funding administered through HPSTAR, the National Natural Science Foundation of China (12488201), and the National Key Research and Development Project of China (2021ZD0301800, 2022YFA1403103).
 \end{acknowledgments}

\bibliography{CB}

\appendix

\begin{figure*}[t]
\includegraphics[angle=0,scale=0.6]{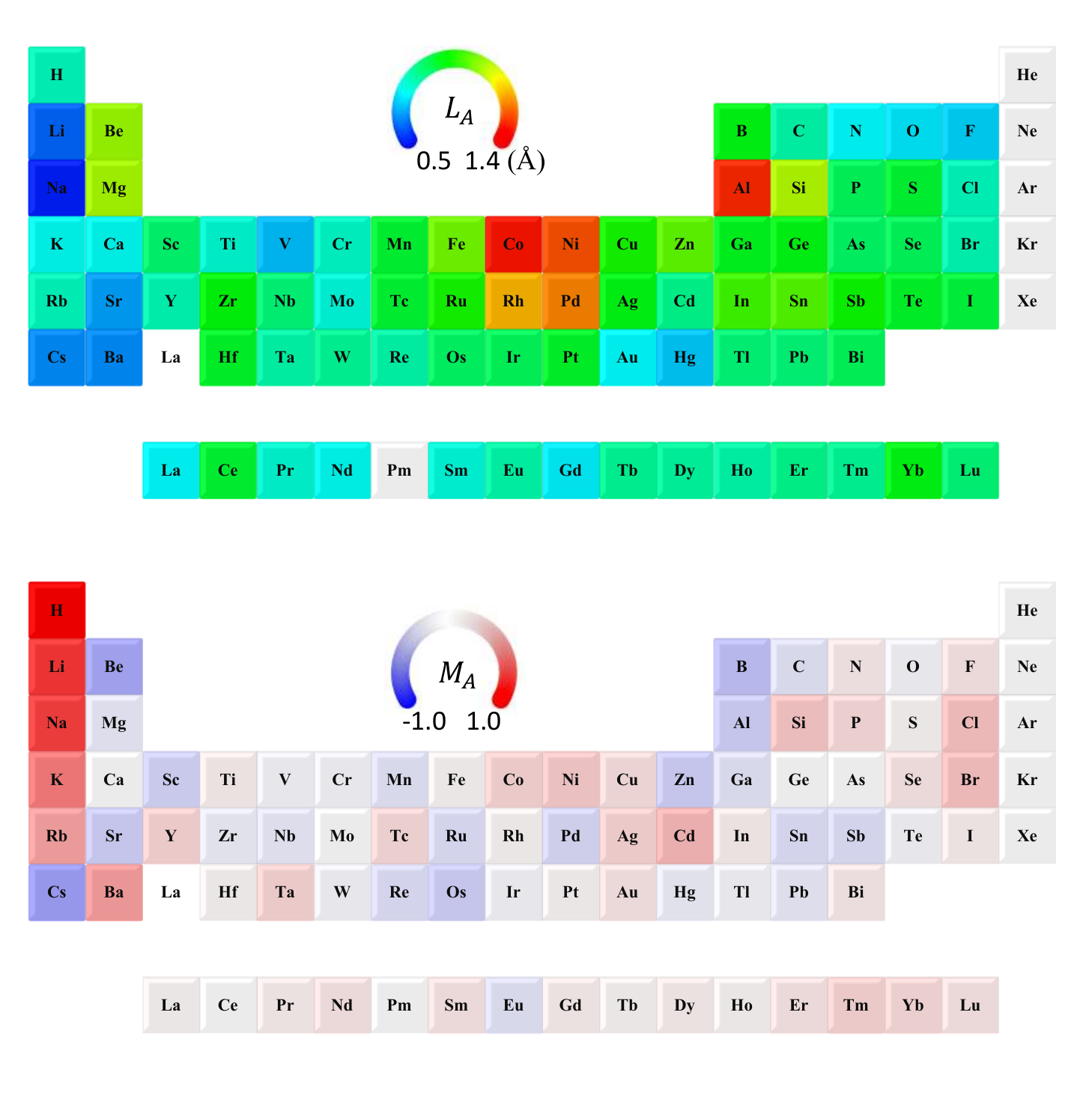}
\caption{
 Visualization of the characteristic decay length $L_A$ and the valence-state modulation factor $M_A$ for Bonding Attractivity.
 }
\label{Fig_LM}
\end{figure*}

\begin{figure*}[t]
\includegraphics[angle=0,scale=0.2]{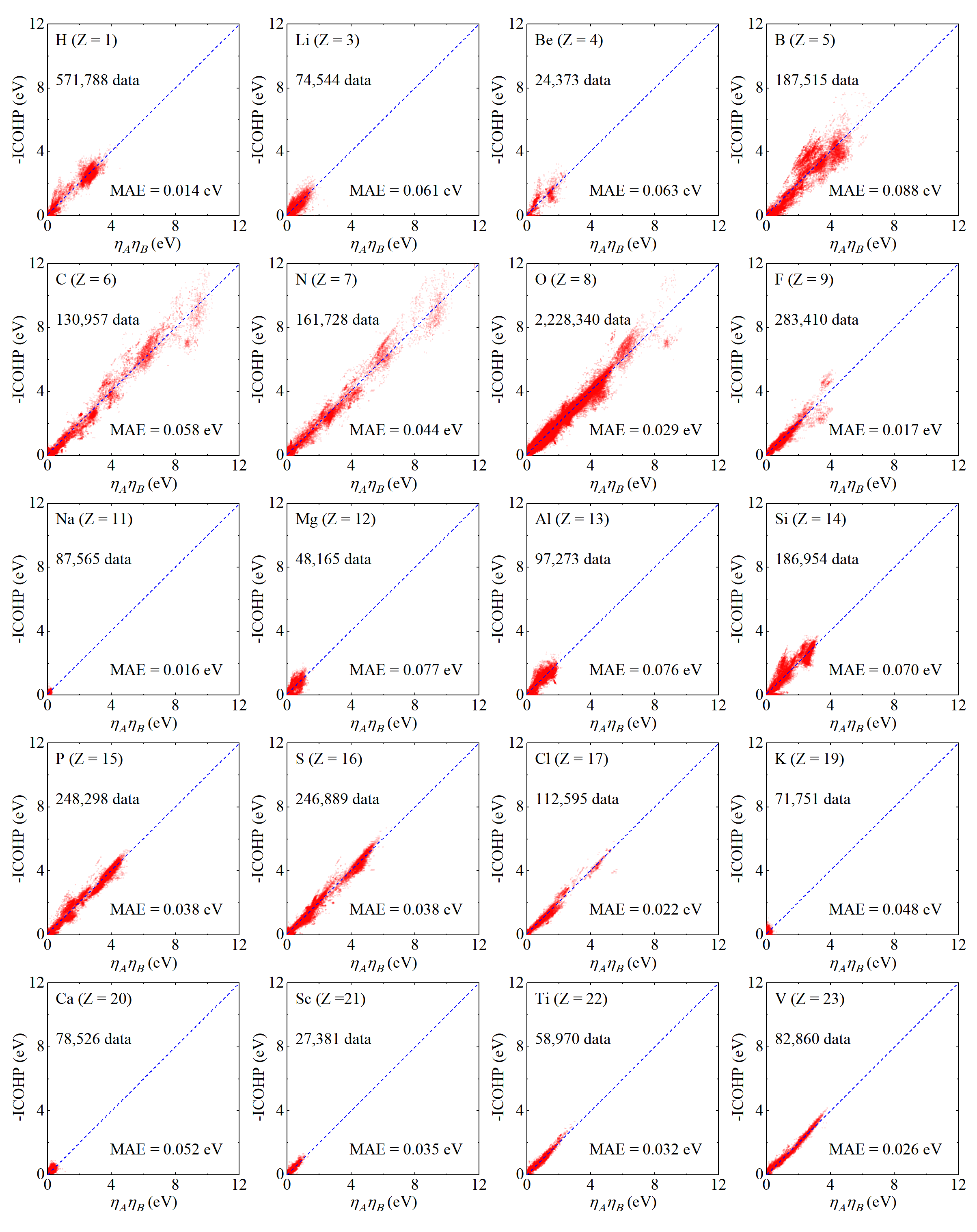}
\caption{
 Parity plots of BA-predicted ICOHP versus DFT-calculated ICOHP, resolved by the central element, from Z = 1 to Z = 23. Each panel aggregates all $A-B$ bonds that include the labeled element, with the diagonal indicating perfect agreement. Numbers inside each panel denote the sample size used for that element’s bonds. 
 }
\label{BA-1}
\end{figure*}

\begin{figure*}[t]
\includegraphics[angle=0,scale=0.2]{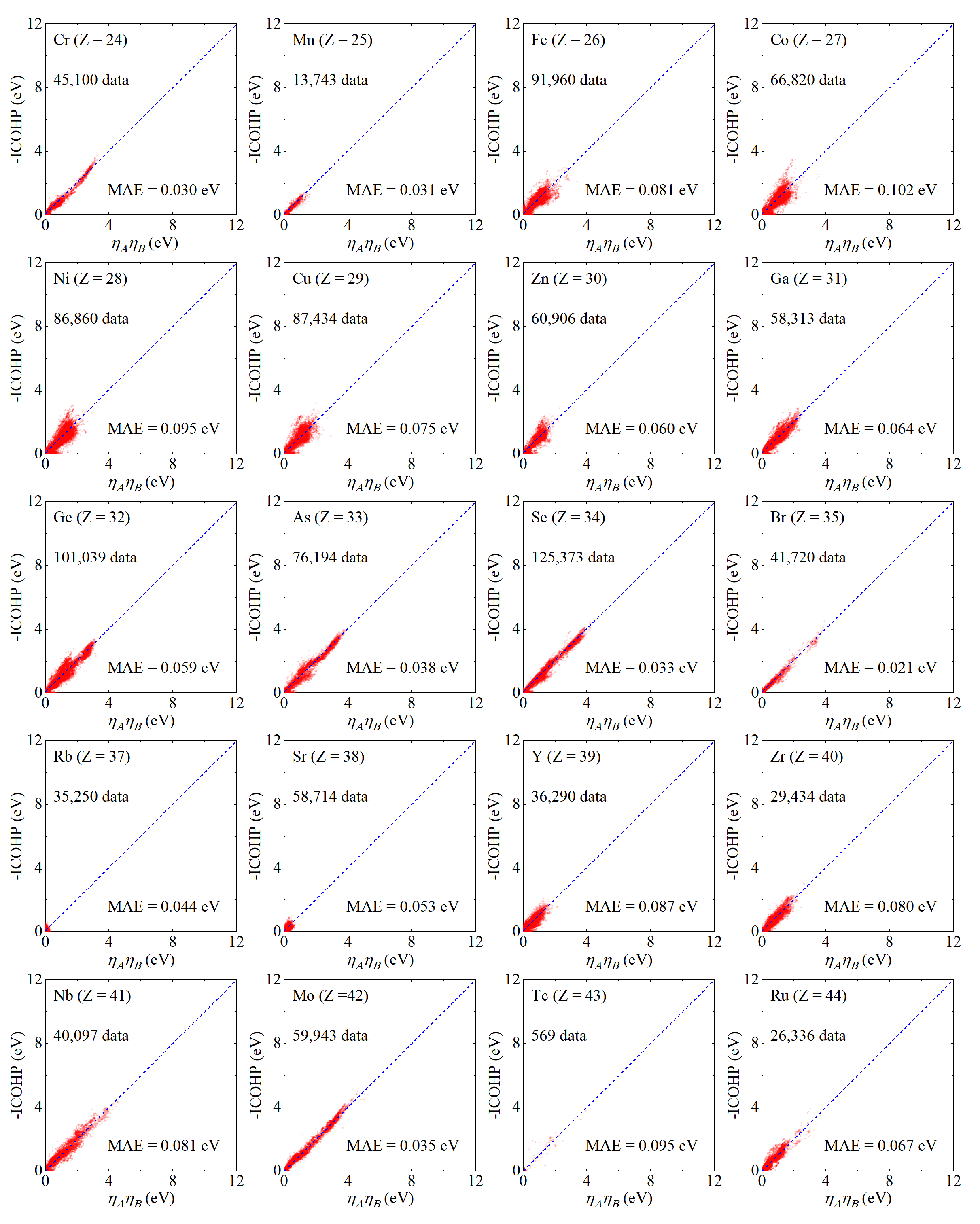}
\caption{
 Parity plots of BA-predicted ICOHP versus DFT-calculated ICOHP, resolved by the central element, from Z = 24 to Z = 44. Each panel aggregates all $A-B$ bonds that include the labeled element, with the diagonal indicating perfect agreement. Numbers inside each panel denote the sample size used for that element’s bonds. 
 }
 \label{BA-2}
\end{figure*}

\begin{figure*}[t]
\includegraphics[angle=0,scale=0.2]{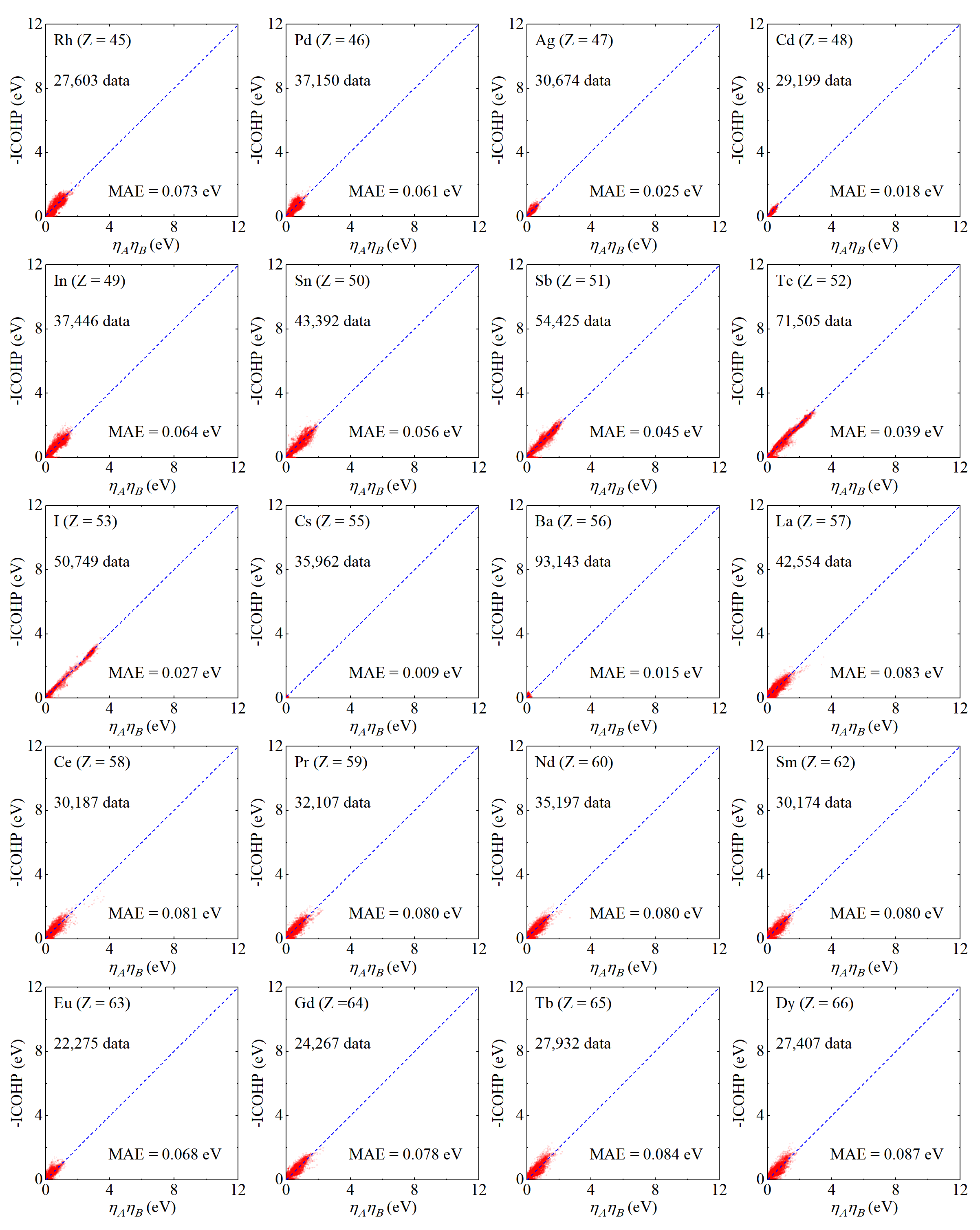}
\caption{
 Parity plots of BA-predicted ICOHP versus DFT-calculated ICOHP, resolved by the central element, from Z = 45 to Z = 66. Each panel aggregates all $A-B$ bonds that include the labeled element, with the diagonal indicating perfect agreement. Numbers inside each panel denote the sample size used for that element’s bonds. 
 }
\label{BA-3}
\end{figure*}

\begin{figure*}[t]
\includegraphics[angle=0,scale=0.2]{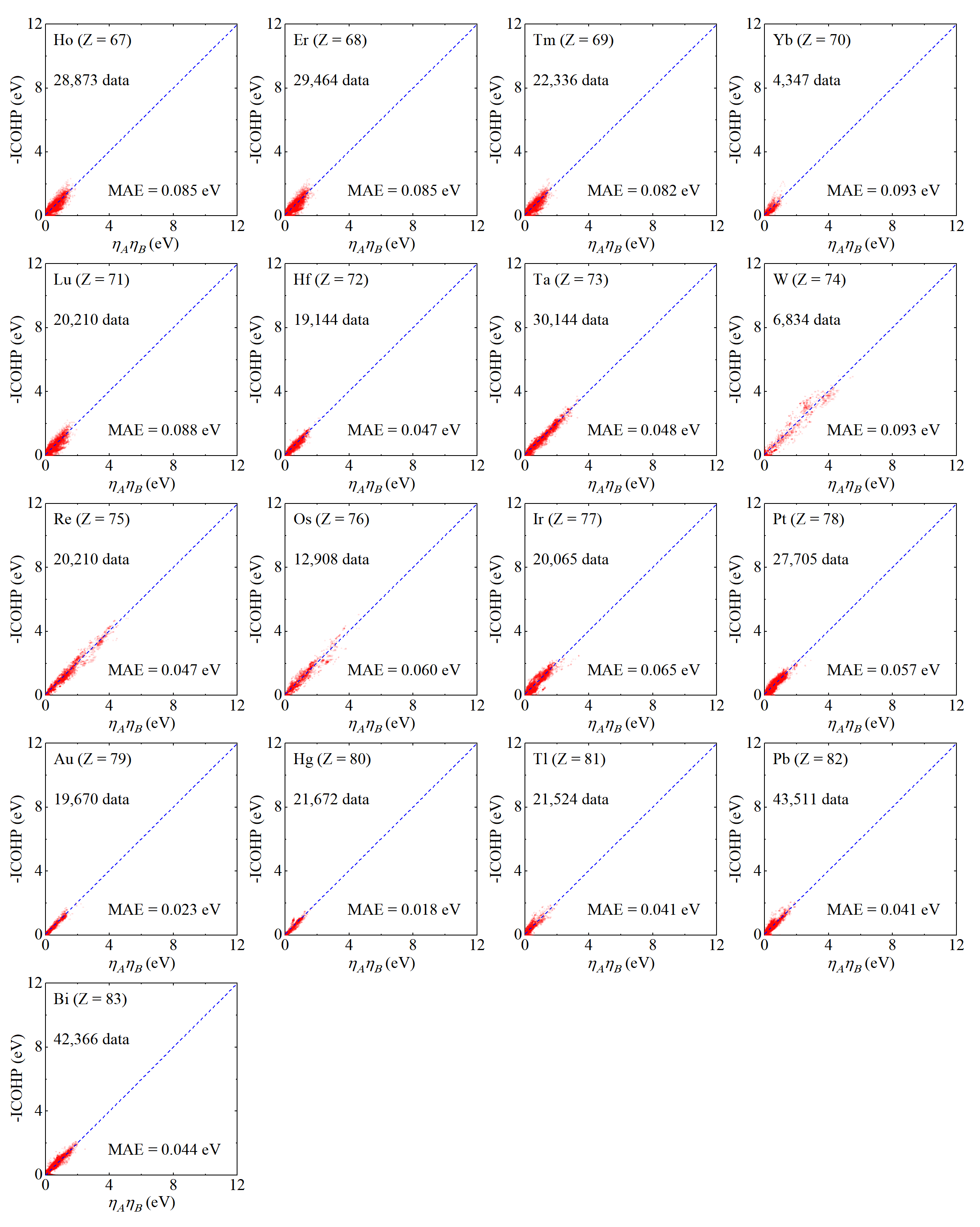}
\caption{
 Parity plots of BA-predicted ICOHP versus DFT-calculated ICOHP, resolved by the central element, from Z = 67 to Z = 83. Each panel aggregates all $A-B$ bonds that include the labeled element, with the diagonal indicating perfect agreement. Numbers inside each panel denote the sample size used for that element’s bonds. 
 }
\label{BA-4}
\end{figure*}

\section{First-principles calculation}

 The electronic structures and atomic-orbital projections of more than 50,000 materials were investigated in a high-throughput manner based on the density functional theory (DFT)~\cite{dft1, dft2} calculations as implemented in the Quantum ESPRESSO (QE) package~\cite{pwscf}. 
 The interactions between electrons and nuclei were described by ultrasoft pseudopotentials as implemented in the PSLIBRARY~\cite{psl}. 
 The generalized gradient approximation (GGA) of Perdew-Burke-Ernzerhof (PBE)~\cite{PBE} type was adopted for the exchange-correlation functional. 
 For elements from Z=1 (H) to Z=83 (Bi) present in our dataset, their valence orbital configurations and the number of atomic orbitals used in the Wannier downfolding are listed in TABLE \ref{tab_ele}. 
 The wave-function and charge-density cutoffs for the plane-wave basis were set to 1.3 times the suggested values by PSLIBRARY. 
 A Gamma-centered $\boldsymbol{k}$-point mesh with a grid spacing of 0.2 \AA$^{-1}$ was adopted for the Brillouin zone (BZ) sampling. 
 The Gaussian smearing method with a width of 0.004 Ry was employed for the Fermi surface broadening. 
 The lattice constants and internal atomic positions were taken directly from the Materials Project database. 
 All calculations were performed within a nonmagnetic framework. 

\begin{table*}[tb]
\caption{
\label{tab_ele}
List of the valence orbital configurations, orbital numbers \cite{psl}, and covalent radii \cite{RA} of all elements used in our calculations and analysis.
}
\begin{center}
\begin{tabular*}{16cm}{@{\extracolsep{\fill}} cccccccc}

\hline \hline
Element & Orbital Conf. & Orbital Num. & $r_A$ (\AA) &  Element & Orbital Conf. & Orbital Num. & $r_A$ (\AA)  \\
\hline
H  & $1s$ & 1 & 0.31 & Tc & $4s4p5s4d5p$ & 13 & 1.47 \\
He & $1s$ & 1 & 0.28 & Ru & $4s4p5s4d$ & 10 & 1.46 \\
Li & $1s2s2p$ & 5 & 1.28 & Rh & $4s4p5s4d$ & 10 & 1.42 \\
Be & $2s2p$ & 4 & 0.96 & Pd  & $4s4p5s4d$ & 10 & 1.39\\
B & $2s2p$ & 4 & 0.84 & Ag  & $4s4p5s4d$ & 10 & 1.45\\
C & $2s2p$ & 4 & 0.76 & Cd  & $4s4p5s4d$ &10 & 1.44 \\
N & $2s2p$ & 4 & 0.71 & In  & $4d5s5p$ & 9 & 1.42 \\
O & $2s2p$ & 4 & 0.66 & Sn  & $4d5s5p$ & 9 & 1.39 \\
F & $2s2p$ & 4 & 0.57 & Sb & $5s5p$ & 4 & 1.39\\
Ne & $2s2p$ & 4 & 0.58 & Te & $5s5p$ & 4 & 1.38\\
Na & $2s2p3s$ & 5 & 1.66 & I & $5s5p$ & 4 & 1.39 \\
Mg & $2s2p3s3p$ & 8 & 1.41 & Xe & $4d5s5p$ & 9 & 1.40 \\
Al & $3s3p$ & 4 & 1.21 & Cs & $5s5p6s$ & 5 & 2.44 \\
Si & $3s3p$ & 4 & 1.11 & Ba & $5s5p6s$ & 5 & 2.15 \\
P & $3s3p$ & 4 & 1.07 & La & $5s5p6s5d6p4f$ & 20 & 2.07 \\
S & $3s3p$ & 4 & 1.05 & Ce & $5s5p6s5d6p$ & 13 & 2.04 \\
Cl & $3s3p$ & 4 & 1.02 & Pr & $5s5p6s5d6p$ & 13 & 2.03 \\
Ar & $3s3p$ & 4 & 1.06 & Nd & $5s5p6s5d6p$ & 13 & 2.01 \\
K & $3s3p4s4p$ & 8 & 2.03 & Pm & $5s5p6s5d6p$ & 13 & 1.99 \\
Ca & $3s3p4s4p$ & 8 & 1.76 & Sm & $5s5p6s5d6p$ & 13 & 1.98 \\
Sc & $3s3p4s3d$ & 10 & 1.70 & Eu & $5s5p6s5d6p$ & 13 & 1.98 \\
Ti & $3s3p4s3d$ & 10 & 1.60 & Gd & $5s5p6s5d6p$ & 13 & 1.96 \\
V & $3s3p4s3d$ & 10 & 1.53 & Tb & $5s5p6s5d6p$ & 13 & 1.94 \\
Cr & $3s3p4s3d$ & 10 & 1.39 & Dy & $5s5p6s5d6p$ & 13 & 1.92 \\
Mn & $3s3p4s3d$ & 10 & 1.39 & Ho & $5s5p6s5d6p$ & 13 & 1.92 \\
Fe & $4s3d4p$ & 9 & 1.32 & Er & $5s5p6s5d6p$ & 13 & 1.89 \\
Co & $4s3d4p$ & 9 & 1.26 & Tm & $5s5p6s5d6p$ & 13 & 1.90 \\
Ni & $4s3d4p$ & 9 & 1.24 & Yb & $5s5p6s5d6p$ & 13 & 1.87 \\
Cu & $4s3d4p$ & 9 & 1.32 & Lu & $5s5p6s5d6p$ & 13 & 1.87 \\
Zn & $4s3d4p$ & 9 & 1.22 & Hf & $5s5p6s5d$ & 10 & 1.75 \\
Ga & $4s3d4p$ & 9 & 1.22 & Ta & $5s5p6s5d$ & 10 & 1.70 \\
Ge & $4s4p$ & 4 & 1.20 & W & $5s5p6s5d6p$ & 13 & 1.62 \\
As & $4s4p$ & 4 & 1.19 & Re & $5s5p6s5d$ & 10 & 1.51 \\
Se & $4s4p$ & 4 & 1.20 & Os & $5s5p6s5d$ & 10 & 1.44 \\
Br & $4s4p$ & 4 & 1.20 & Ir & $5s5p6s5d$ & 10 & 1.41 \\
Kr & $4s3d4p$ & 9 & 1.16 & Pt & $5s5p6s5d$ & 10 & 1.36 \\
Rb & $4s4p5s5p$ & 8 & 2.20 & Au & $5s5p6s5d$ & 10 & 1.36 \\
Sr & $4s4p5s5p$ & 8 & 1.95 & Hg & $5s5p6s5d$ & 10 & 1.32 \\
Y & $4s4p5s4d5p$ & 13 & 1.90 & Tl & $5d6p6s$ & 9 & 1.45 \\
Zr & $4s4p5s4d5p$ & 13 & 1.75 & Pb & $5d6p6s$ & 9 & 1.46 \\
Nb & $4s4p5s4d5p$ & 13 & 1.64 & Bi & $5d6p6s$ & 9 & 1.48 \\
Mo & $4s4p5s4d$ & 10 & 1.54 &  &  & & \\
\hline\hline
\end{tabular*}
\end{center}
\end{table*}

\section{Closest Wannier Function}
  In plane-wave-based DFT calculations, the atomic orbitals used for post-processing and projection analysis are typically incomplete and non-orthogonal:
  \begin{equation}
  \begin{gathered}
  \sum_{a b}\langle n \mid a\rangle S_{a b}^{-1}\langle b \mid n\rangle \neq 1 \\
  S_{a b}=\langle a \mid b\rangle \neq \delta_{a b}.
  \end{gathered}
  \end{equation}
  Here $|n\rangle$ is the calculated Kohn–Sham eigenstate in a plane-wave basis.  
  $|a\rangle$ and  $|b\rangle$ are two atomic orbitals from two neighboring atoms, and $S_{a b}$ is their orbital overlap integral. 
  $\widehat{S}^{-1}$ is the inverse of the overlap matrix $\widehat{S}$. 
  Notably, these incomplete and non-orthogonal atomic orbitals pose significant challenges for bonding analysis.
  Although several strategies have been proposed to address this issue in orthogonal bases (like Lowdin orthogonalization or directly cutting the orbital wave-function in real-space), the incomplete atomic-orbital basis still cannot adequately describe all electronic states of interest, especially for electronic states around or above the Fermi level. 
  An ideal strategy is to construct a set of complete and orthogonal Wannier Functions. 
  Nowadays, the most widely used orbital-downfolding method is based on maximally localized Wannier functions (MLWFs)\cite{mlwf}. 
  However, the construction of MLWFs is highly sensitive to parameter choices. 
  It inevitably involves manual parameter adjustment and is therefore not well suited for high-throughput calculations. 
  In this work, for constructing a set of ideal Wannier functions to represent the original atomic orbital, we adopted a recently developed Closest Wannier Function (CWF) method\cite{cwf}. 
  CWFs are characterized by weak parameter dependence and require no human intervention. 
  Unlike MLWF construction, which requires iterative calculations to optimize an artificial spread function $\Omega$, CWFs are obtained through a single-step singular value decomposition of the projection matrix $\widehat{A}$:
  \begin{equation}
  \widehat{A}=\widehat{U} \widehat{\Sigma} \widehat{V}^{\dagger}.
  \end{equation}
  Here $\widehat{A}$ is a non-square matrix whose elements $A_{a n}$ reflect the projection amplitudes between atomic orbital $|a\rangle$ and the eigen- Kohn-Sham state $|n\rangle$ from DFT calculation. 
  In general, $\widehat{A}$ is not unitary. 
  But we can further define a semi-unitary matrix $\widehat{B}$ using: 
  \begin{equation}
  \widehat{B}=\widehat{U}\widehat{V}^{\dagger}.
  \end{equation}
  Utilizing the half-unitarity of $\widehat{B}$, we can therefore construct a set of orthogonal and complete Wannier function: 
  \begin{equation}
  \left|a_{\mathrm{CWF}}\right\rangle=\sum_n|n\rangle B_{a n}.
  \end{equation}
  As proved in Ref. \cite{cwf}, this set of $\left|a_{\mathrm{CWF}}\right\rangle$ has the minimal deviation from the original guiding functions defined by $\widehat{A}$, and is therefore called the set of Closest WFs.

  The only parameter dependence of the CWF construction lies in the definition of the guiding matrix $\widehat{A}$. 
  In earlier works, this was defined by the atomic orbital projection matrix $\langle a \mid n\rangle$ with a weight of $w_n$:
  \begin{equation}
  A_{a n}=\langle a \mid n\rangle w_n.
  \end{equation}
  The role of $w_n$ is to emphasize the energy window of interest, similar to the frozen windows used in MLWF. 
  Here we let 
  \begin{equation}
  w_n=\frac{1}{e^{\frac{\varepsilon_n-\varepsilon_f-\Delta}{\Delta}}+1}+o
  \end{equation}
  to ensure that all occupied states can be well described by CWFs.  
  $\Delta$ is a broadening parameter to control the energy window above the Fermi level. 
  $o$ is a small finite quantity introduced to ensure that all singular values of $\widehat{A}$ are non-zero. 
  In this work, we set $\Delta$ = 2.0 eV and $o$ = 0.01. 
  
  In addition to $w_n$, in the definition of guiding matrix $\widehat{A}$, we also introduced the dual basis. 
  Because of the nonorthogonality of the original set of atomic orbitals $|a\rangle$, we can define a set of corresponding dual basis functions, $|a'\rangle$: 
  \begin{equation}
  |a'\rangle=\sum_b|b\rangle S_{b a}^{-1}.
  \end{equation}
  In mathematical terms, the dual orbital $|a'\rangle$ describes the unique dual vector associated with $|a\rangle$ and is therefore orthogonal to the other atomic orbitals:
  \begin{equation}\langle a' \mid b\rangle=\delta_{a b}.\end{equation}
  Here we further introduced two projection matrices $\widehat{C}$ and $\widehat{D}$ for the original atomic orbital $|a\rangle$ and its dual orbital $|a'\rangle$, respectively: 
  \begin{equation}\begin{array}{l}
  C_{a n}=\langle a \mid n\rangle \\
  D_{a n}=\langle a' \mid n\rangle.
  \end{array}\end{equation}
  In practice, the types and number of orbitals included in the CWF downfolding can significantly influence the bonding analysis.
  Typically, certain outer-shell orbitals contribute minimally to atomic bonding because they are nearly unoccupied, such as the Mg $3p$ orbitals or the Ti $4p$ orbitals. 
  However, due to their spatial extension, they can easily overlap with other atomic orbitals and capture electronic states that do not physically belong to them, like the core electronic state. 
  To avoid undue dependence on these insignificant outer-shell orbitals, we used a composite strategy when constructing the $\widehat{A}$ matrix: 
  \begin{equation}
  A_{a n}=[f_n D_{a n}+\left(1-f_n\right) C_{a n}] w_n .
  \end{equation}
  Here $f_n$ is the Fermi distribution of the eigenstate $|n\rangle$. 
  As shown by this expression, we used the projection matrix of the dual orbital, $D_{a n}$, for occupied states. 
  This can effectively reduce the influence of outer-shell orbitals on the bond analysis of occupied states. 

  In the MLWF downfolding calculation\cite{mlwf}, one needs to optimize a local indicator $\Omega$. 
  This involves the overlap calculation between the wave-functions of different $\boldsymbol{k}$ points: $\left\langle m\boldsymbol{k}\right|e^{-i\boldsymbol{br}}|n\boldsymbol{k+b}\rangle$.
  But for CWFs, the calculations at different $\boldsymbol{k}$-points are fully independent. 
  We just need to calculate the CWF transition matrix $\widehat{B}(\boldsymbol{k})$ at each $\boldsymbol{k}$-point. 
  The final real-space tight-binding model and RDM in the CWF basis are obtained by Fourier transformation of $H(\boldsymbol{k})_{ab}$ and $D(\boldsymbol{k})_{ab}$, respectively: 
  \begin{equation}
  \begin{gathered}
  H(\boldsymbol{R})_{ab}=\sum_{\boldsymbol{k}} e^{-i \boldsymbol{k}\boldsymbol{R}} H(\boldsymbol{k})_{ab} \\
  H(\boldsymbol{k})_{ab}=\sum_n \varepsilon_{n \boldsymbol{k}} B_{a n}(\boldsymbol{k}) B_{b n}^*(\boldsymbol{k})
  \end{gathered}
  \end{equation}
  \begin{equation}
  \begin{gathered}
  D(\boldsymbol{R})_{ab}=\sum_{\boldsymbol{k}} e^{-i \boldsymbol{k}\boldsymbol{R}} D(\boldsymbol{k})_{ab} \\
  D(\boldsymbol{k})_{ab}=\sum_n f_{n \boldsymbol{k}} B_{a n}(\boldsymbol{k}) B_{b n}^*(\boldsymbol{k})
  \end{gathered}
  \end{equation}

  Notably, because CWFs are constructed to be as close as possible to the guiding functions, they are not necessarily well localized at the same time like the MLWF. 
  Thus, within a finite $\boldsymbol{k}$-mesh or under Born–von Karman (BvK) boundary conditions, sometimes the long-distance orbital hopping cannot fully decay to zero. 
  Usually, this may slightly influence the effectiveness of Wannier interpolation. 
  However, for the short-distance bonds between adjacent atoms of interest, their orbital hybridization can still be well described. 

\section{Bond analysis of ICOHP}
  In this work, we employed the CWF-based integrated crystal orbital Hamilton population (ICOHP) method to quantitatively describe atomic bond strength \cite{cohp0,cohp,cohp2}. 
  Crystal Orbital Hamilton Population (COHP) is a computational tool used in quantum chemistry and solid-state physics to analyze the electronic structure of molecules and solids. 
  Within the DFT framework, ICOHP represents the contribution of orbital hybridization and helps elucidate the covalent character of chemical bonds in materials. 
  
  After CWF downfolding, we can rewrite the electronic eigenwavefunctions $|n \boldsymbol{k}\rangle$ in the CWF basis: 
  \begin{eqnarray}
  |n \boldsymbol{k}\rangle=\sum_{Aa} C_{Aa}^{n \boldsymbol{k}}|A a \boldsymbol{k}\rangle,
  \label{eq_coopcohp}
  \end{eqnarray}
  Here $|A a \boldsymbol{k}\rangle$ denotes the CWF of atomic orbital $a$ on atom $A$ in $\boldsymbol{k}$-space, and $C_{Aa}^{n \boldsymbol{k}}$ is the corresponding combination coefficient.  
  The band structure $\varepsilon_{n \boldsymbol{k}}$ can therefore be written as: 
  \begin{equation}
  \begin{aligned}
  \label{cohp_nk}
  \varepsilon_{n \boldsymbol{k}} & =\langle n \boldsymbol{k}| \widehat{H}|n \boldsymbol{k}\rangle=\sum_{Aa,Bb} C_{B b}^{n \boldsymbol{k} *} C_{A a}^{n \boldsymbol{k}}\langle B b \boldsymbol{k}| \widehat{H}|A a \boldsymbol{k}\rangle \\
  & =\sum_{\boldsymbol{R} A a, \boldsymbol{R}^{\prime} B b} C_{B b}^{n \boldsymbol{k} *} C_{A a}^{n \boldsymbol{k}} e^{i \boldsymbol{k}\left(\boldsymbol{R}^{\prime}-\boldsymbol{R}\right)}\left\langle\boldsymbol{R}^{\prime} B b\right| \widehat{H}|\boldsymbol{R} A a\rangle.
  \end{aligned}
  \end{equation}
  Here we employed a Fourier transform $|A a \boldsymbol{k}\rangle=\sum_R e^{-i \boldsymbol{k} \boldsymbol{R}}|\boldsymbol{R} A a\rangle$ for the bond analysis between specific atomic pairs in real-space. 
  In the summation of Eq.~\ref{cohp_nk}, the contribution from the orbital pair $|\boldsymbol{R} A a\rangle$ and $|\boldsymbol{R}^{\prime} B b\rangle$ provides the corresponding COHP in the $|n \boldsymbol{k}\rangle$ state:
  \begin{equation}
  \operatorname{COHP}^{n \boldsymbol{k}}_{\boldsymbol{R} A a, \boldsymbol{R}^{\prime} B b}=C_{B b}^{n \boldsymbol{k} *} C_{A a}^{n \boldsymbol{k}} e^{i \boldsymbol{k}\left(\boldsymbol{R}^{\prime}-\boldsymbol{R}\right)}\left\langle\boldsymbol{R}^{\prime} B b\right| \widehat{H}|\boldsymbol{R} A a\rangle.
  \end{equation}  
  In general, one is often more concerned with its distribution in energy space: 
  \begin{equation}
  \operatorname{COHP}_{\boldsymbol{R} A a, \boldsymbol{R}^{\prime} B b}(\varepsilon)=\sum_{n \boldsymbol{k}} \delta\left(\varepsilon-\varepsilon_{n \boldsymbol{k}}\right)  \operatorname{COHP}^{n \boldsymbol{k}}_{\boldsymbol{R} A a, \boldsymbol{R}^{\prime} B b}
  \end{equation}
  The ICOHP between the orbitals $|\boldsymbol{R} A a\rangle$ and $|\boldsymbol{R}^{\prime} B b\rangle$ is defined by integrating $\operatorname{COHP}_{\boldsymbol{R} A a, \boldsymbol{R}^{\prime} B b}(\varepsilon)$ below the Fermi level:
  \begin{equation}
  \begin{gathered}
  \operatorname{ICOHP}_{\boldsymbol{R} A a, \boldsymbol{R}^{\prime} B b}=\int_{-\infty}^{\varepsilon_f} d \varepsilon \operatorname{COHP}_{\boldsymbol{R} A a, \boldsymbol{R}^{\prime} B b}(\varepsilon) \\
  =\langle \boldsymbol{R} A a| \widehat{D}\left|\boldsymbol{R}^{\prime} B b\right\rangle\left\langle \boldsymbol{R}^{\prime} B b\right| \widehat{H}|\boldsymbol{R} A a\rangle
  \end{gathered}
  \end{equation}
  Here we introduce the reduced density matrix (RDM) operator $\widehat{D}$: 
  \begin{equation}
  \widehat{D}=\sum_{n \boldsymbol{k}}|n \boldsymbol{k}\rangle f_{n \boldsymbol{k}}\langle n \boldsymbol{k}|,
  \end{equation}
  whose off-diagonal terms also carry information about bond order. $f_{n \boldsymbol{k}}$ is the Fermi–Dirac occupation with a finite-temperature broadening of 0.2 eV. The total ICOHP for the atomic pair $(\boldsymbol{R} A,\boldsymbol{R}^{\prime} B)$ requires an additional summation of their hybrid orbitals:
  \begin{equation}
  \operatorname{ICOHP}_{\boldsymbol{R} A, \boldsymbol{R}^{\prime} B}=\sum_{a \in A, b \in B} \operatorname{ICOHP}_{\boldsymbol{R} A a, \boldsymbol{R}^{\prime} B b}
  \end{equation}

\end{document}